# DO THERMOELECTRIC GENERATOR MODULES DEGRADE DUE TO NICKEL DIFFUSION

P.V.Gorskyi[1,2], DSc (phys-math), gena.grim@gmail.com

[1)]Institute of Thermoelectricity of the NAS and MES of Ukraine, [2)]Yu.Fedkovych Chernivtsi National University

*The paper shows by calculation that the diffusion of nickel even for 50 years does not lead to degradation of thermoelectric generator modules. In the process, we used the theory of composites to calculate the electrical contact resistance, our own diffusion theory of electrical contact resistance, as well as the method for approximating the temperature dependences of thermoelectric material characteristics from the experimental data. When using the above method, it was assumed that the main mechanism of scattering of free charge carriers in a thermoelectric material is their scattering on the deformation potential of acoustic phonons with a free path length independent of energy but inversely proportional to temperature, and the main mechanism of phonon scattering is phonon-phonon scattering with Umklapp, which is not affected by the nickel impurity in the thermoelectric material. Thus, it was believed that the role of nickel is reduced only to a change in the concentration of free charge carriers in the material.*

*Introduction. The current state of the problem, the purpose and objectives of the research*. Contact structures in thermoelectric modules of "standard" size with a thermoelectric leg length of 1.5-3 mm are created by soldering. In order to prevent the diffusion of solder components into thermoelectric material (TEM), which is believed to lead to undesirable changes in the parameters and characteristics of TEM, an anti-diffusion layer is deposited on the surface by galvanic or purely chemical means. The material of this layer can be mainly nickel or cobalt and less often iron or manganese. Nevertheless, among specialists in the field of thermoelectricity, there are certain concerns that the diffusion of these materials in TEM will lead to undesirable changes in its parameters. To dispel these fears, the authors of [1] experimentally studied the transport properties and thermoelectric figure of merit of the $Cu_{0.01}Bi_2Te_{2.7}Se_{0.3}$ alloy doped with nickel, cobalt, iron, and manganese in the amount of 2 at.%. It was found that despite the change in concentration in the range $(2.8 \div 5.3) \cdot 10^{19}$ cm$^{-3}$ in the temperature range from 5 to 350 K, the thermoelectric figure of merit practically does not depend on the doping impurity, and in a wider range, an impurity, for example, nickel, reduces the thermoelectric figure of merit only up to 475 K, and then increases it. Based on the analysis of the obtained data, the authors concluded that in the investigated temperature range, the diffusion of the material of the anti-diffusion layer in TEM does not lead to module degradation. But since the diffusion process itself was not studied by the authors their results do not yet give an idea of how exactly TEM, and, consequently, the thermoelectric generator module, will behave for a long period of time. The study of this issue is the purpose of this article.

*Solution of the equation of non-stationary diffusion of metal in TEM and its consequences.*

If the diffusion coefficient $D$ of metal in TEM is considered constant, then in a one-dimensional approximation this equation will be given by:

$$\frac{\partial c}{\partial t} = D \frac{\partial^2 c}{\partial x^2}, \qquad (1)$$

where $c$ is concentration of metal atoms, $t$ is time, $x$ is coordinate counted from the interface deep into TEM.

For the case of a semi-confined medium, the solution of Eq.(1) can be taken in the form:

$$c = c_0 \operatorname{erfc}\left(x/\sqrt{2Dt}\right), \qquad (2)$$

where $\operatorname{erfc}(...)$ is the so-called additional error integral. Solution (2) satisfies the obvious initial condition $c(t=0)=0$ and the obvious boundary condition $c(x=0)=c_0$. In [2], based on Eq.(2) and the condition of matter balance, it was shown that the growth of the transient layer occurs according to the law $x_0 = 13.552\sqrt{Dt_0}$. Thus, the thickness $h$ of the spent part of nickel layer is matched by the thickness $6.771\sqrt{\pi}h \approx 12h$ of transient contact layer. In this case, the concentration distribution of metal atoms in the transient layer, normalized to its thickness, is determined by the formula:

$$c(x) = c_0 \operatorname{erfc}(6.771\, x/x_0), \qquad (3)$$

and it is shown in Fig.1.

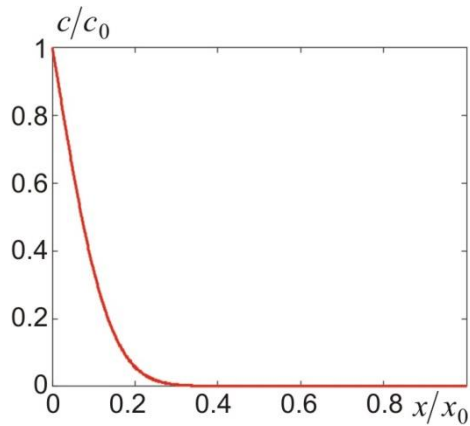

*Fig.1. Distribution of metal atoms in the transient contact layer*

It is noteworthy that the ratio between the thickness of the transient contact layer and the thickness of the spent nickel layer is an invariant of the diffusion process that determines the formation of the transient contact layer. The value of the diffusion coefficient determines only the time dependence of the thickness of the transient contact layer and, consequently, the contact resistance of the TEM-metal.

After depletion of the nickel surface layer, its distribution in TEM is described by the relation:

$$c(x,t) = \frac{1}{\sqrt{\pi D t}} \int_0^{12h} \mathrm{erfc}\left(\frac{y}{\sqrt{\pi} h}\right) \exp\left[-\frac{(x-y)^2}{4Dt}\right] dy. \qquad (4)$$

The distribution of nickel in the transient layer for different moments of time in the process of prolonged diffusion after the depletion of the transient layer is shown in Fig. 2.

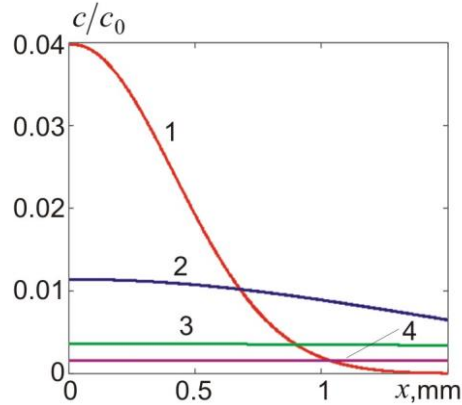

*Fig. 2. Distributions of nickel concentration in a leg 1.5 mm long with a thickness of the original nickel layer of 5 μm after: 1 – 1 month; 2 – 1 year; 3 – 10 years; 4 – 50 years*

It can be seen from Fig. 2 that the leveling of the concentration in a leg 1.5 mm long occurs approximately 50 years after the depletion of the original nickel layer. It should be noted that the diffusion coefficients of nickel in *p*- and *n*-type materials differ by orders of magnitude [3]. In the calculation, we took the larger of the values, namely $3.1 \cdot 10^{-14}$ cm$^2$/s, as the diffusion coefficient *D*. From Fig. 2 and data [4] on the atomic concentration of nickel, it follows that the concentration of nickel atoms in TEM after leveling will be $3.644 \cdot 10^{19}$ см$^{-3}$. Assuming that in the original TEM at 300 K, the thermoEMF is 160 μV/K, the scattering index is r=–0.5, and the effective mass of free charge carriers is $m^* = m_0$, it turns out that the concentration of carriers in it is $3.47 \cdot 10^{19}$ cm$^{-3}$. Thus, in the course of leveling the nickel concentration in TEM, a kind of "doping" of TEM occurs, as a result of which, after 50 years, the concentration of charge carriers in TEM will increase by about 2.05 times, if we assume that each nickel atom supplies one free carrier to the impurity TEM charge of the same sign. Touching on the comparison, although not entirely correct, of this result with the results of work [1], we note that in the case of nickel doping, their concentration increased by a factor of 1.54 compared to the original TEM.

*Calculation of temperature dependences of electrical and thermal contact resistances.* Let us determine these dependences using the theory of composites and taking into account the phenomenon of percolation associated with the formation of metal clusters in the thickness of the transient layer. To do this, we first determine the volume fraction of nickel in the transient contact layer:

$$v(x) = \frac{(A_m/\rho_m)\text{erfc}(6.771x)}{(A_m/\rho_m)\text{erfc}(6.771x) + (M_s/\rho_s)\text{erf}(6.771x)}, \qquad (5)$$

where $A_m$, $\rho_m$ are atomic mass and metal density, respectively, $M_s$, $\rho_s$ are molecular mass and TEM density, respectively, following which the coordinate dependence of specific electrical conductivity of transient contact layer will be determined as:

$$\sigma(x) = 0.25\left\{2\sigma_s - \sigma_m + 3v(x)(\sigma_m - \sigma_s) + \sqrt{[2\sigma_s - \sigma_m + 3v(x)(\sigma_m - \sigma_s)]^2 + 8\sigma_m\sigma_s}\right\}, \qquad (6)$$

where $\sigma_s, \sigma_m$ are the conductivities of TEM and metal at an arbitrary temperature *T*, and, therefore, the electrical contact resistance of TEM-metal at an arbitrary temperature *T* can be determined as follows:

$$r_c(T) = 12h\int_0^1 \frac{dx}{\sigma(x)}, \qquad (7)$$

where *h* is the thickness of spent nickel layer.

The electrical conductivity of metal will be considered to be inversely proportional to temperature:

$$\sigma_m = \sigma_{m0} \cdot (T_0/T), \qquad (8)$$

where $\sigma_{m0}$ is the known value of electrical conductivity of metal at a certain temperature $T_0$.

The temperature dependence of TEM electrical conductivity is found in the following order. First, knowing its thermoEMF $\alpha_{s0}$ at the same temperature $T_0$ and taking into account the above determined value of the scattering index, from the relation

$$\alpha_{s0} = \frac{k}{e}\left(\frac{2F_1(\eta_0)}{F_0(\eta_0)} - \eta_0\right) \qquad (9)$$

we find the combined chemical potential of the subsystem of free charge carriers at this temperature. Further, taking into account the condition of constancy of the concentration of free charge carriers and neglecting the temperature dependence of their effective mass, from the equation

$$\left(\frac{T}{T_0}\right)^{1.5}\frac{F_{1/2}(\eta)}{F_{1/2}(\eta_0)} - 1 = 0, \qquad (10)$$

where $F_m(y)$ are the Fermi integrals determined by the relation

$$F_m(y) = \int_0^\infty \frac{dx}{\exp(x-y)+1}, \qquad (11)$$

we find the temperature dependence of the combined chemical potential $\eta$. After that, considering the constants of the material, which determine the scattering of charge carriers on the deformation potential of acoustic phonons, independent of temperature and knowing the specific electrical conductivity of TEM $\sigma_{s0}$ at temperature $T_0$, we will find its electrical conductivity at an arbitrary temperature:

$$\sigma_s = \sigma_{s0}\left(\frac{T_0}{T}\right)^{1.5}\frac{F_{1/2}(\eta_0)F_0(\eta)}{F_{1/2}(\eta)F_0(\eta_0)}. \qquad (12)$$

The results of calculating the temperature dependences of the electrical contact resistance for a couple of bismuth telluride - nickel with the thickness of the spent nickel layer equal to 5 and 20 µm, respectively, are shown in Fig.3.

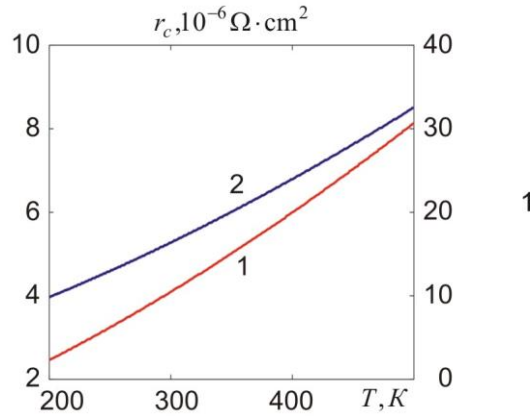

Fig.3. Temperature dependences of contact resistance of bismuth telluride-nickel couple at the thicknesses of spent layer equal to: 1 – 5 µm (left axis), 2 – 20 µm (right axis)

When constructing the plots, the following electrical conductivity values of nickel and bismuth telluride were taken at $T_0$=300K: $\sigma_{m0} = 1.43\cdot 10^5$ Cm/cm, $\sigma_{s0} = 1.4\cdot 10^3$ Cm/cm, and, besides, $A_m = 58$, $\rho_m = 9100\,\text{kg/m}^3$, $M_s = 802$, $\rho_s = 7860$ kg/m$^3$. From the figure it is seen that at the thickness of spent nickel layer 5 µm the electrical contact resistance in the considered temperature range varies from $2.5\cdot 10^{-6}$ to $8\cdot 10^{-6}$ Ohm·cm$^2$, and at the thickness of spent nickel layer 20 µm — from $10^{-5}$ to $3.2\cdot 10^{-5}$ Ohm·cm$^2$. Traditionally, the anti-diffusion layer is made 20 µm thick. Therefore, a comparison of the calculated values of the electrical contact resistance with those traditionally "assigned" and used in the design of thermoelectric generators and coolers allows us to conclude that the thickness of the diffusing nickel layer does not exceed 5 µm, while the rest of 15 µm remain on the surface.

The thermal conductivity of transient contact layer is determined as:

$$\kappa(x) = 0.25\left\{2\kappa_s - \kappa_m + 3\nu(x)(\kappa_m - \kappa_s) + \sqrt{[2\kappa_s - \kappa_m + 3\nu(x)(\kappa_m - \kappa_s)]^2 + 8\kappa_m\kappa_s}\right\}, \qquad (13)$$

and, hence, its specific thermal contact resistance is equal to:

$$r_t(T) = 12h \int_0^1 \frac{dx}{\kappa(x)}. \qquad (14)$$

In so doing, the thermal conductivity of metal $\kappa_m$ will be considered temperature independent, and the thermal conductivity of TEM $\kappa_s$ is found as follows:

$$\kappa_s = L\sigma_s T + (\kappa_{s0} - L_0 \sigma_{s0} T_0)\frac{T_0}{T}. \qquad (15)$$

The Lorentz numbers in formula (15) are found in the following way:

$$L = \left(\frac{k}{e}\right)^2 \left[\frac{3F_2(\eta)}{F_0(\eta)} - 4\frac{F_1(\eta)^2}{F_0(\eta)^2}\right], \qquad (16)$$

$$L_0 = \left(\frac{k}{e}\right)^2 \left[\frac{3F_2(\eta_0)}{F_0(\eta_0)} - 4\frac{F_1(\eta_0)^2}{F_0(\eta_0)^2}\right]. \qquad (17)$$

When deriving formula (15), it was assumed that the thermal conductivity of TEM due to free charge carriers is subject to the Wiedemann-Franz law, and its lattice thermal conductivity is subject to the Leibfried-Schlemann law.

The temperature dependences of specific thermal contact resistance of bismuth telluride – nickel couple are shown in Fig.4.

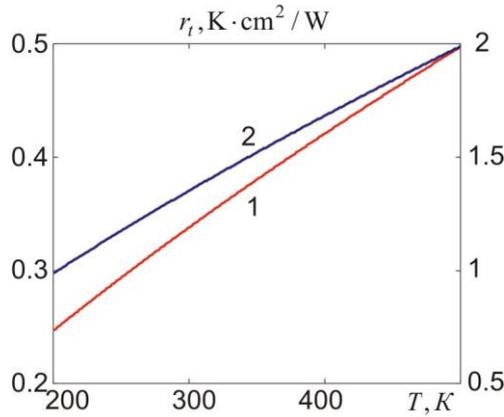

*Fig.4. Temperature dependences of thermal contact resistance of bismuth telluride-nickel couple at the thicknesses of spent layer equal to: 1 – 5 μm (left axis), 2 – 20 μm (right axis)*

When constructing the plots, the following thermal conductivity values of nickel and TEM were taken at $T_0$=300K: $\kappa_{m0} = 0.92 \text{ W/(cm·K)}$, $\kappa_{s0} = 1.7 \cdot 10^{-2} \text{ W/(cm·K)}$.

From Fig.3 it is seen that in the temperature range of 200 – 500 K the specific thermal resistance of transient layer at the thickness of spent nickel layer 5 μm varies from 0.25 to 0.5 K·cm²/W, and at the thickness of spent nickel layer 20 μm – from 1 to 2 K·cm²/W.

The thermoEMF of transient layer will be found as:

$$\alpha = \frac{\int_0^1 [\alpha_m \kappa_m^{-1} v(x) + \alpha_s \kappa_s^{-1}(1 - v(x))] dx}{\int_0^1 [\kappa_m^{-1} v(x) + \kappa_s^{-1}(1 - v(x))] dx}, \qquad (18)$$

where $\alpha_m$ is thermoEMF of metal, $\alpha_s$ is thermoEMF of TEM, which is found by the formula:

$$\alpha_s = \frac{k}{e}\left(\frac{2F_1(\eta)}{F_0(\eta)} - \eta\right) \qquad (19)$$

Figs.5 – 7 show the temperature dependences of thermoEMF, power factor and dimensionless thermoelectric figure of merit of contact layer.

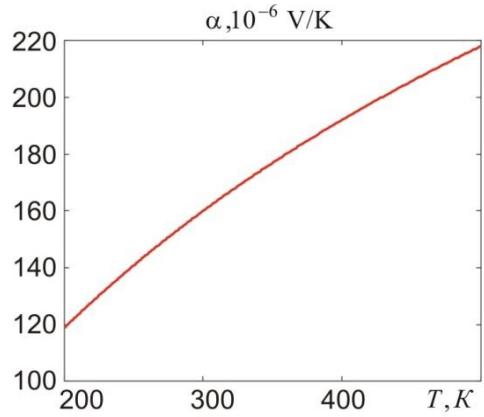

Fig.5. Temperature dependence of thermoEMF of transient contact layer

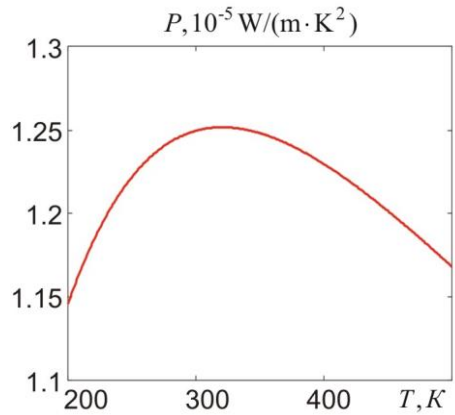

*Fig.6. Temperature dependence of power factor of transient contact layer*

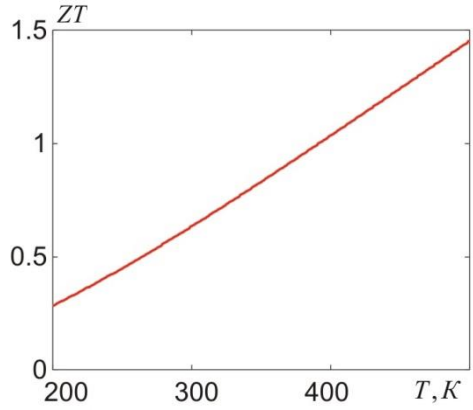

*Fig.7. Temperature dependence of thermoelectric figure of merit of transient contact layer*

From the figures it is seen that in the temperature range of 200-500 K, the thermoEMF of transient contact layer increases from 120 to 220 μV/K, the power factor is maximum at a temperature about 320 K, and reaches $1.25 \cdot 10^{-5}$ W/(m·K$^2$), and the dimensionless thermoelectric figure of merit grows from 0.3 to 1.5. This growth is due to the fact that the transient contact layer is considered not as a doped material, but as a composite, the components of which, when included in the composite, retain their thermoelectric characteristics, including temperature dependences. This feature distinguishes composites from doped thermoelectric materials, in which the role of doping impurities consists either in supplying free charge carriers of one type or another, or in changing phonon scattering mechanisms.

We have already seen that complete leveling of the nickel concentration in a leg 1.5 mm long, even at the highest diffusion coefficient inherent in an *n*-type material, occurs no earlier than after 50 years, and even then, provided that the entire leg is at a temperature of 500 K, which is not the case in real working conditions.

In so doing, the equilibrium concentration of nickel in the thermoelectric leg is comparable with the concentration of free charge carriers in the initial TEM. Therefore, it is of some interest to consider the effect of such a peculiar "doping" on the dimensionless thermoelectric figure of merit of TEM.

We carry out this review in the following order. We will assume that the influence of "doping" is reduced to the supply of free charge carriers, and each nickel atom supplies one carrier. First, we find the concentration of free charge carriers in the original TEM from the ratio:

$$n_0 = \frac{4(2\pi m^* k T_0)^{1.5} F_{1/2}(\eta_0)}{\sqrt{\pi} h^3} . \qquad (20)$$

Assuming that $m^* = m_0$ and taking into account the above results, we find that $n_0 = 3.47 \cdot 10^{19}$ cm$^{-3}$, and, therefore, the relative increase in the concentration of free charge carriers after leveling will be δ=1.05.

The temperature dependence of the chemical potential of the subsystem of free charge carriers in the "doped" TEM can be found from the equation:

$$\frac{F_{1/2}(\eta)T^{1.5}}{F_{1/2}(\eta_0)T_0^{1.5}(1+\delta)} - 1 = 0. \quad (21)$$

The specific electrical conductivity of the "doped" TEM will be found from the relation:

$$\sigma_s = \sigma_{s0}(1+\delta)\left(\frac{T_0}{T}\right)^{1.5}\frac{F_{1/2}(\eta_0)F_0(\eta)}{F_{1/2}(\eta)F_0(\eta_0)}. \quad (22)$$

After that, using relations (15) – (17) and (19), we will find the thermal conductivity, thermoEMF, and ultimately – the dimensionless thermoelectric figure of merit of the "doped" material. The results of such a calculation are shown in Fig. 8.

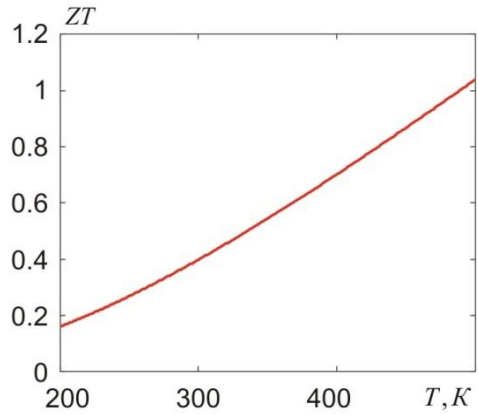

*Fig.8. Temperature dependence of the dimensionless thermoelectric figure of merit of doped TEM*

Based on this dependence, one can estimate the efficiency of a thermoelement in the electric power generation mode. In the maximum power mode, this efficiency is equal to

$$\eta_{\max P} = \frac{T_h - T_c}{T_h} \cdot \frac{1}{4/Z_m T_h + 2 - (T_h - T_c)/T_h}, \quad (23)$$

where $Z_m$ is the average thermoelectric figure of merit in the range from $T_c$ to $T_h$, and the value of 6.2% is obtained for it.

In the mode of maximum efficiency it is equal to:

$$\eta_{\max P} = \frac{(T_h - T_c)\left[\sqrt{1+0.5(ZT_h + ZT_c)} - 1\right]}{T_h\left[\sqrt{1+0.5(ZT_h + ZT_c)} + T_c/T_h\right]}, \quad (24)$$

and the value of 6.5% is obtained for it. Thus, we come to the conclusion that nickel diffusion, even for 50 years, does not adversely affect the operation of the thermoelement.

The predicted values of efficiency can be reduced, in particular, by the temperature dependence of the charge carrier concentration and by the temperature dependence of thermal conductivity, which differs from the Leibfried-Schlemann law.

*Conclusions*

1. It is shown that the diffusion of nickel at the initial stage, when the thickness of its spent layer does not exceed 5 – 20 µm, does not have a negative effect on the electrical and thermal contact resistances of the TEM-metal. In the temperature range of 200-500 K and the thicknesses of spent nickel layer 5 – 20 µm the electrical contact resistance varies in the interval from $2.5 \cdot 10^{-6}$ to $3.2 \cdot 10^{-5}$ Ohm·cm$^2$, the thermal contact resistance – in the interval from 0.25 to 2 K·cm$^2$/W, the thermoEMF – in the interval from 120 to 220 µV/K. The power factor, varying in the interval $(1.15–1.25) \cdot 10^{-5}$ W/(m·K$^2$), reaches maximum at 320 K, and the dimensionless thermoelectric figure of merit of transient layer, if considered as a composite, reaches 1.5 at 500 K. Thus, at the initial stage, TEM with nickel that has diffused in it can be considered as a functionally graded TEM.

2. It is shown that leveling of nickel concentration in TEM even for 50 years does not adversely affect the operation of the thermoelement in the electric power generation mode, since even after this period in the temperature range of 200 – 500 K its dimensionless thermoelectric figure of merit changes from 0.18 to 1.03, which between extreme temperatures of 300 and 500 K provides an efficiency of 6.2% in the highest power mode and 6.5% in the maximum efficiency mode.